\documentclass[conference]{IEEEtran}
\IEEEoverridecommandlockouts
\usepackage{cite}
\usepackage{amsmath,amssymb,amsfonts}
\usepackage{algorithmic}
\usepackage{graphicx}
\usepackage{tabularx}
\usepackage{ragged2e}
\usepackage{textcomp}
\usepackage{xcolor}
\usepackage{caption}
\usepackage{url}
\usepackage{subcaption}

\def\BibTeX{{\rm B\kern-.05em{\sc i\kern-.025em b}\kern-.08em
    T\kern-.1667em\lower.7ex\hbox{E}\kern-.125emX}}
\graphicspath{{./figure/}}
\begin{document}

\title{Partially Detected Intelligent Traffic Signal Control:  Environmental Adaptation
}

\author{
    \IEEEauthorblockN{Rusheng Zhang\IEEEauthorrefmark{1}, Romain Leteurtre\IEEEauthorrefmark{2}, Benjamin Striner\IEEEauthorrefmark{3},
    Ammar Alanazi \IEEEauthorrefmark{4}, 
    Abdullah Alghafis \IEEEauthorrefmark{4},
    Ozan K. Tonguz\IEEEauthorrefmark{1}}
    \IEEEauthorblockA{\IEEEauthorrefmark{1}Department of Electrical and Computer Engineering, Carnegie Mellon University, Pittsburgh, USA
    \\rushengz@andrew.cmu.edu, tonguz@ece.cmu.edu}
    \IEEEauthorblockA{\IEEEauthorrefmark{2}Faculty of Communication Systems, Ecole Polytechnique F{\'e}d{\'e}rale de Lausanne, Lausanne, Switzerland
    \\romain.leteurtre@epfl.ch}
    \IEEEauthorblockA{\IEEEauthorrefmark{3}Department of Machine Learning, Carnegie Mellon University, Pittsburgh, USA
    \\bstriner@cs.cmu.edu}
    \IEEEauthorblockA{\IEEEauthorrefmark{4}King Abdulaziz City for Science and Technology, Riyadh, Saudi Arabia
    \\\{salanazi,alghafis\}@kacst.edu.sa}
}
\maketitle

\begin{abstract}
\textit{Partially Detected Intelligent Traffic Signal Control} (PD-ITSC) systems that can optimize traffic signals based on limited detected information could be a cost-efficient solution for mitigating traffic congestion in the future. 
In this paper, we focus on a particular problem in PD-ITSC -- adaptation to changing environments. To this end, we investigate different reinforcement learning algorithms, including Q-Learning, Proximal Policy Optimization (PPO), Advantage Actor-Critic (A2C), and Actor-Critic with Kronecker-Factored Trust-Region (ACKTR).
Our findings suggest that RL algorithms can find optimal strategies under partial vehicle detection; however, policy-based algorithms can adapt to changing environments more efficiently than value-based algorithms. We use these findings to draw conclusions about the value of different models for PD-ITSC systems. 
\end{abstract}


\section{Introduction}
\footnote[1]{The research reported in this paper was partially funded by King Abdulaziz City for Science and Technology (KACST), Riyadh, Kingdom of Saudi Arabia}
Intelligent Traffic Signal Control (ITSC) systems have attracted the attention of researchers and the general public as a means of alleviating traffic congestion. ITSC systems specifically attempt to make informed decisions about traffic signals using real-time vehicle information from a variety of detectors \cite{lowrie1990scats, hunt1982scoot,gartner1983opac}. However, these traditional ITSC systems are not financially viable for large-scale ITSC deployments. In this paper, we focus on Partially Detected Intelligent Traffic Signal Control (PD-ITSC), a new form of ITSC that has recently received increasing attention from both academia and industry  \cite{zhang2018partially,zhang2018increasing,tonguz2019harness}. Such systems operate even if only a subset of all vehicles are detected and use partial information to control traffic signals accordingly. An example scenario involving partial detection would be partial adoption of vehicle-mounted devices such as Dedicated Short-Ranged Communications (DSRC) radios or cellphones. Therefore, PD-ITSC is a technology of great interest since it can benefit from the rapid growth of the Internet of Things (IoT) and mobile computing.

 Most traditional ITSC systems operate with an ideal detection scheme, assuming that every vehicle is observed by the agent controlling the traffic signal, and \textit{are not optimized for partial vehicle detection}.
In contrast to fully-observable systems, PD-ITSC systems can be deployed with partial technology penetration (e.g. 10-20\% of vehicles are equipped with a communication device) and adapt to changing detection rates over time. 
 Reinforcement Learning (RL) is a powerful tool for this scenario as it does not require a comprehensive theoretical modeling of the environment. Our previous work has established the viability of RL for PD-ITSC using Q-learning in a static environment \cite{zhang2018partially}. In this paper, we consider several other RL algorithms and evaluate performance in a dynamically changing environment.




Specifically, we investigate several RL algorithms including Q-Learning, Proximal Policy Optimization (PPO), Advantage Actor-Critic (A2C), and Actor-Critic with Kronecker-Factored Trust-Region (ACKTR), and provide quantitative comparisons of the performance of these different RL algorithms.

As a key result, we find that all the RL algorithms investigated in this paper can find near-optimal strategies under partial detection of vehicles, but policy-based algorithms can adapt to changing environments better than value based algorithms.  

\section{Background}\label{background}
\subsection{Reinforcement Learning in Traffic Control}

Recent developments in Deep Reinforcement Learning (DRL) have generated a huge interest in applying DRL to traffic control \cite{yau2017survey}. Several research groups have applied Deep Q-learning (DQL) to traffic control, while others have evaluated the potential of using Q-learning for the traffic signal control problem \cite{van2016deep, genders2016using, van2016coordinated}. Research parallel to DQL has also proven the viability of several policy-based methods, such as policy gradient \cite{mousavi2017traffic} and Deep Deterministic Policy Gradient (DDPG) \cite{casas2017deep}. 

The research described above relies on traditional ITSC where \textit{all the vehicles are detected}. The use cases of such systems are limited since well-established mathematical solutions have been available for decades. In this paper, we apply the RL strategy to a new form of traffic signaling systems, PD-ITSC systems, which we believe will be a fundamental component of future traffic control applications.

\subsection{Partially Detected Intelligent Traffic Signal Control}
\label{ss:pdts}
The rapid development of the Internet of Things (IoT) in the last 5 years has brought forth new wireless technologies applicable to vehicle sensing for ITS. They include DSRC, C-V2X, RFID, Bluetooth, Ultra-Wide Band (UWB), Zigbee, and even cellphone apps such as Google Map \cite{ friesen2015bluetooth, qu2010intelligent, tonguzred, zhang2018virtual, zhang2019using}. All these systems are more cost-efficient than traditional loop detectors or cameras. Performance-wise,
 these systems are able to provide finer-grained information than traditional systems but only for equipped vehicles.

\begin{figure}[ht]
\centering
\includegraphics[width=2.5in]{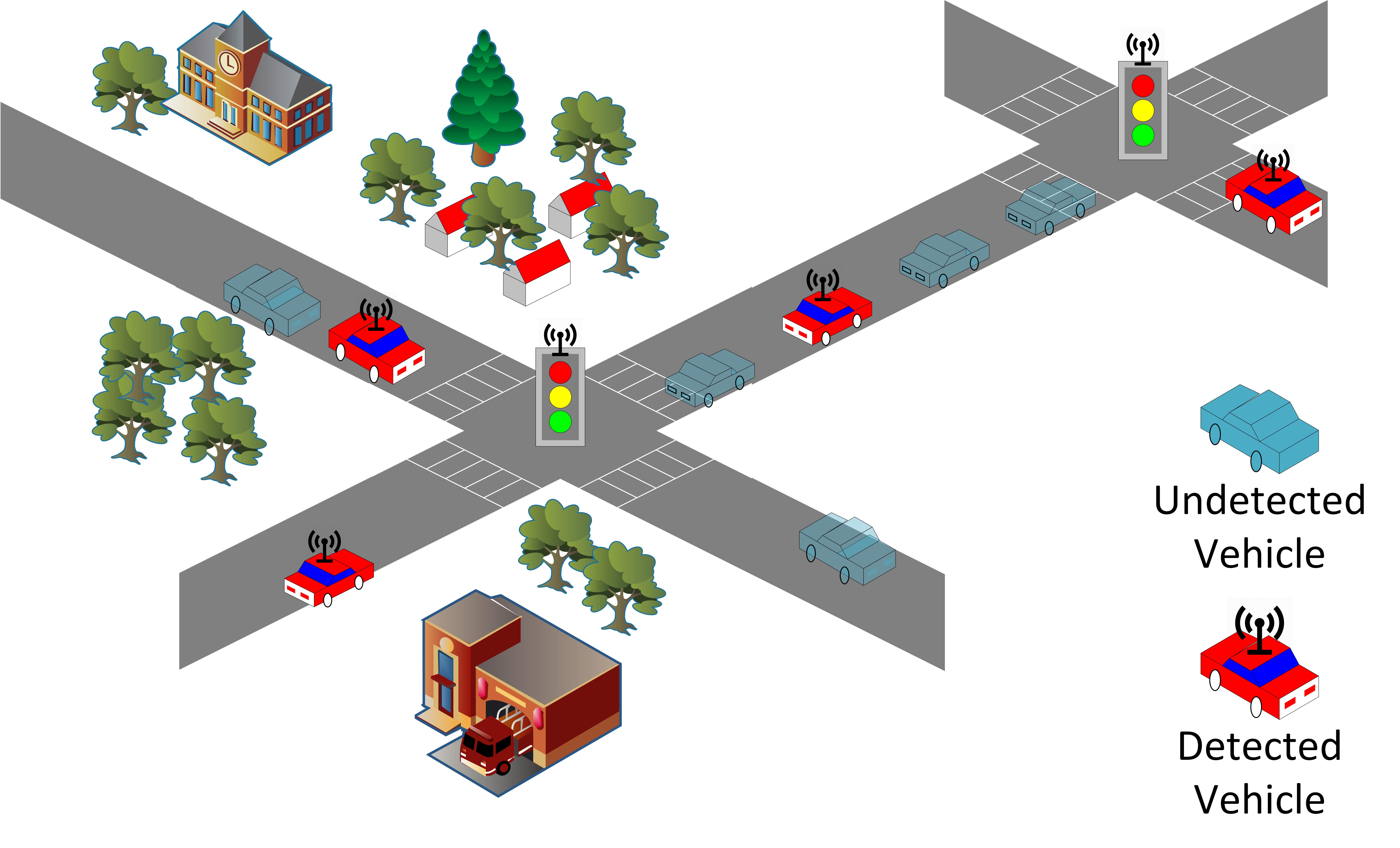}
\caption{Illustration of a PD-ITSC system. }
\label{fig_PDTS}
\end{figure}

 In a PD-ITSC system, both equipped and unequipped  vehicles co-exist in the traffic network. Based on the detected vehicles information, the traffic lights decide on the current phase at the intersections, to minimize the overall waiting time for both detected vehicles and undetected vehicles, as illustrated in Figure \ref{fig_PDTS}.


\begin{figure}[ht]
\centering
\includegraphics[width=3in]{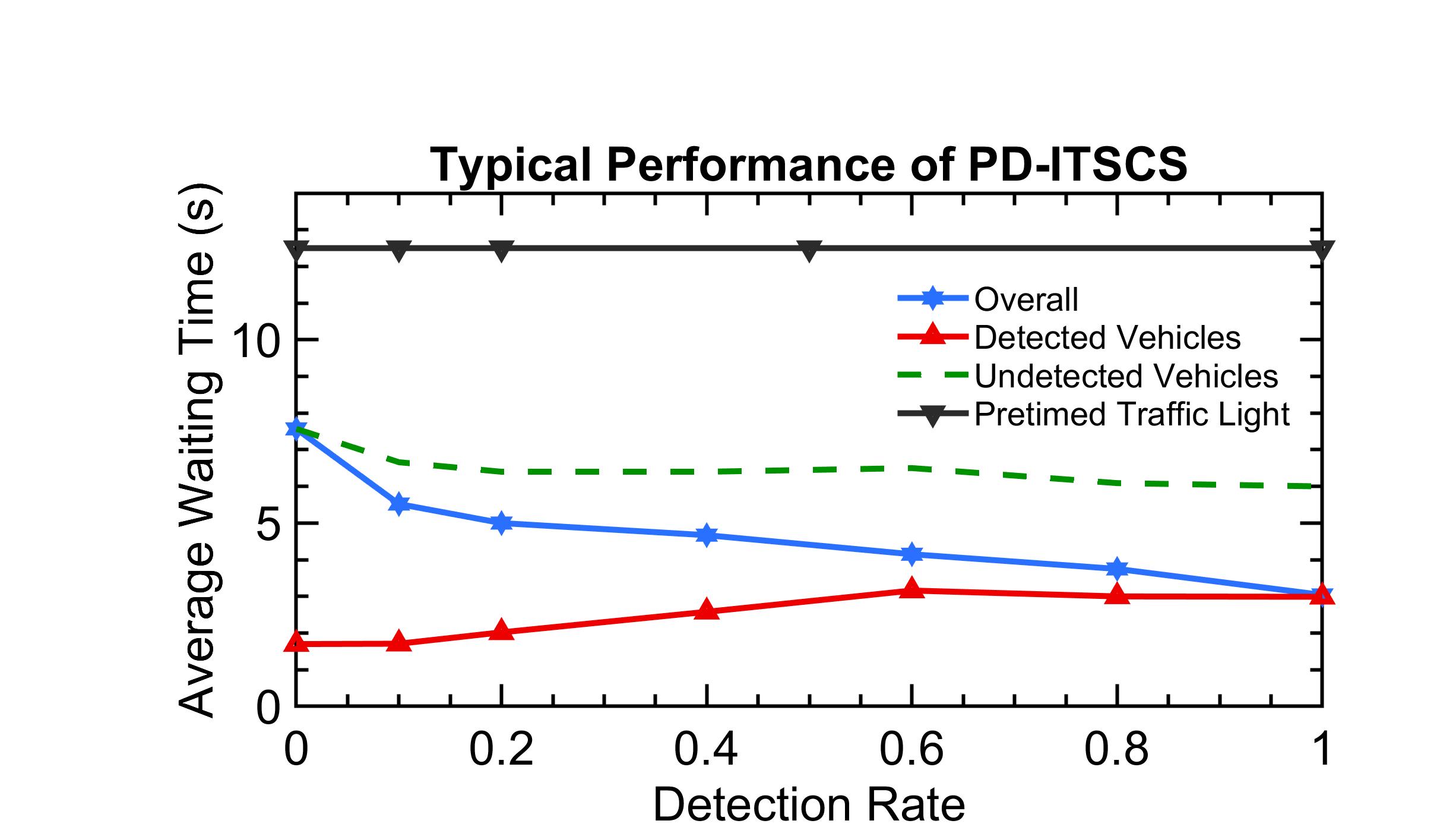}
\caption{Increasing detection rate decreases waiting time in a typical PD-ITSC system (results from \protect\cite{zhang2018partially})}
\label{fig_ill}
\end{figure}

Previous studies have shown that a signal control scheme based on Q-learning is capable of handling varying car flow rates (light and heavy traffic) and varying detection rates, showing that RL could be a promising solution for PD-ITSC \cite{zhang2018partially}. Figure \ref{fig_ill} shows the typical performance of a PD-ITSC system trained with Q-learning. 

PD-ITSC provides a viable adoption strategy. The high price associated with traditional ITSC is borne by the Department of Transportation to install detectors on road surfaces or nearby. In PD-ITSC, detected vehicles have a shorter average waiting time than undetected vehicles, incentivising users to equip their vehicles with a communication device, and companies can profit by selling and installing such devices. It is financially feasible to equip most of the intersections of a city with PD-ITSC, which would not be feasible with a traditional ITSC.

\section{Problem Statement}\label{problem}

During initial deployment of a PD-ITSC system, a small percentage of cars might be equipped and, therefore, detected. As time goes by, more commuters will recognize the benefit of being detected (shown in Figure \ref{fig_ill}) and gradually adopt the devices or services to equip their vehicles. An agent should be able to update itself to the change in detection rate over time. One solution is to periodically update the agent manually. This would be very cumbersome and require constant maintenance and expense, especially for scenarios involving  large-scale deployment. Our goal in this paper is to present a solution that can adapt to changing detection rates without manual maintenance and re-calibration.


\section{Methodology}\label{method}
\subsection{Reinforcement Learning Algorithms}
The goal of reinforcement learning is to train an agent that interacts with the environment by selecting an action in a way that maximizes cumulative reward. At every time step, the agent gets an observation of the state and reward information from the environment and chooses an action. During this process, the agent tries to maximize the cumulative reward. In this paper, we apply several RL algorithms to PD-ITSC, including Q-Learning, PPO, A2C, and ACKTR \cite{watkins1992q, mnih2016asynchronous,schulman2017proximal,wu2017scalable}.

\subsubsection{Q-Learning}

In the Q-learning approach \cite{watkins1992q}, the agent learns the state-action value known as 'Q-Value', $Q(s_t, a_t)$, which is the expected cumulative discounted future reward given state and action. The cumulative discounted future reward is defined as:
\begin{equation}
    Q(s_t,a_t) = r_t + \gamma r_{t+1} + \gamma^2 r_{t+2} + \gamma^3 r_{t+3}+...
\end{equation}

Here, $r_t$ is the reward at each time step, the meaning of which needs to be specified according to the actual problem, and $\gamma<1$ is the discount factor. At every time step, the agent updates its Q function by an update of the Q value:
\begin{equation}
    Q(s_t,a_t) = Q(s_t,a_t) + \alpha(r_{t+1}+\gamma \max{Q(s_{t+1},a_t)}-Q(s_t,a_t))
\end{equation}

\subsubsection{A2C}
A2C algorithm is a policy-based algorithm \cite{mnih2016asynchronous}, which tries to learn a probabilistic policy $\pi_\theta(a_t|s_t)$, which is the probability of taking a certain action in certain state, $\theta$ is the policy parameter. The algorithm is considered to have two parts: an actor which determine the agent's action; a critic that tells the actor how good the action is. At each update, the policy parameter $\theta$ can be updated by:
\begin{equation}
    \theta \gets \theta + \alpha \nabla_\theta(\log \pi_\theta(s_t,a_t))A(s_t,a_t)
\end{equation}

Where $\alpha$ is the learning rate,  $A(s_t,a_t)$ is the advantage function, the value can be approximated by the value function: 
\begin{equation}
    A(s_t,a_t) =R_t \gamma V_\omega(s_{t+1})-V(s_t)
\end{equation}
$V_\omega$ is the average reward of the state, known as critic, and $\omega$ is the critic model parameter. The agent can learn the value function by:
\begin{equation}
    \omega \gets \omega + \beta \nabla_\omega (R_t+\gamma V_\omega(s_{t+1})-V_\omega(s_t))V_\omega (s_t,a_t)
\end{equation}
where $\beta$ is a different learning rate for learning the value function.

\subsubsection{PPO}
PPO algorithm is based on Actor-critic and Trust Region Policy Optimization (TRPO)\cite{schulman2017proximal}.  Unlike TRPO, which uses a standalone constraint, namely, the trust region to constrain the update size \cite{schulman2015trust}, PPO achieves the similar goal by using the clipped surrogate objective:
\begin{equation}
    L^{CLIP}(\theta) = \hat{\mathbb E}[\min(r_t(\theta)\hat{A_t}, \operatorname{clip}(r_t(\theta), 1-\epsilon, 1+\epsilon)\hat{A_t})]
\end{equation}
where $r_t(\theta) = \frac{\pi_\theta(a_t|s_t)}{\pi_{\theta_k}(a_t|s_t)}$ denotes how close the sample policy is to the current policy. In this way, the objective function is clipped if the two policies deviate too much, and hence prevent big destructive step size in policy update.

\subsubsection{ACKTR}
The ACKTR algorithm is a modified version of the TRPO method \cite{wu2017scalable}. The ACKTR algorithm applies the Kronecker-factored approximation to optimize both actor and critic. The ACKTR method uses Kronecker-Factored Approximate Curvature (K-FAC) as the approximation of the Fisher Information Matrix (FIM) to reduce computational complexity of TRPO.  KFAC calculates small blocks of a matrix layer by layer:
\begin{equation}
    F_l \approx  \mathbb E [aa^T]\otimes \mathbb E[\nabla_sL(\nabla_sL)^T] := A\otimes S := \hat F_l
\end{equation}
where $a$ is activate input of the layer, $L$ is the log likelihood of the current policy, and $s$ is the pre-activation output of the layer,
where $\nabla_WL = (\nabla_sL)a^T$, $W$ is the weight of a certain layer, $L$ is the output of the layer and $a$ is the activate input of the layer. The weight $W$ of each layer becomes:
\begin{equation}
    \Delta W = \hat F^{-1}vec\{\nabla_W L\}
\end{equation}
In this way, the computational complexity is reduced to the order of $W$. To apply K-FAC method to both the actor and critic, in a shared model, one can consider the output as a joint distribution of two independent distributions, the action-state distribution (actor) and the value distribution (critic). 

\subsection{Parameter Modeling}
\label{SS:para}

\subsubsection{Agent action}

In our context, the relevant action of the agent is either to keep the current traffic light phase or to switch to the next traffic light phase. At every time step, the agent makes an observation and takes an action accordingly, achieving intelligent control of traffic.

\subsubsection{Reward}
\label{sss:reward}
As a primary concern of this paper, the goal is to decrease the average delay of commuters in the network, by using traffic light phasing strategy $\mathcal{S}$. We want to find the best traffic light phasing strategy $\mathcal{S}$, such that $t_\mathcal{S} - t_{\min}$ is minimized, where $t_\mathcal{S}$ is the average travel time of commuters in the network, under the traffic control scheme $\mathcal{S}$, and $t_{\min}$ is the physically possible lowest average travel time. It has been shown in  previous work \cite{zhang2018partially} that maximizing the following reward function at each time $t$ minimizes the delay for all commuters:
\begin{equation}
R_t = -\sum_{c\in C} \frac{1}{v_{\max, c}}[v_{\max,c}-v_{\mathcal{S},c}(t)]
\label{eq:reward}
\end{equation}
where $R_t$ denotes immediate reward, $v_{\max, c}$ is the maximum vehicle speed $c$, $v_{\mathcal{S},c}(t)$ is the vehicle speed under strategy $\mathcal{S}$ of vehicle $c$ and $C$ is the set of all vehicles on the street. The reward function described in (\ref{eq:reward}) is known as \emph{full reward} since the summation is over all vehicles on the street. However, the full reward can't be directly observed in the real world by the agent, since the system is not able to perceive undetected vehicles. 

To overcome this difficulty, in this paper, we choose \emph{partial reward} instead:

\begin{equation}
R^{partial}_t = -\sum_{c\in C_{obs}} \frac{1}{v_{\max, c}}[v_{\max,c}-v_{\mathcal{S},c}(t)]
\label{eq:partialReward}
\end{equation}
where $C_{obs}$ denotes the set of vehicles that can be observed by the traffic control agent. 


\subsubsection{State Representation}
\label{sss:state_repr}


Previously, Compact State Representation (CSR)  was proposed  in \cite{zhang2018partially}, which is shown in Table \ref{tab:stateRep}. This state representation captures the essential information for traffic control and encodes it into a compact vector. The advantage of such representation is its size, as it requires much smaller computational power to work with and hence accelerates the convergence of the algorithm used. In real world deployment, less computation power can reduce the financial cost, and faster convergence rate makes agents adapt to the changing environment much faster.

\begin{table}[ht]
    \centering
    \caption{Compact state representation}
    \label{tab:stateRep}
\begin{tabularx}{.95\linewidth}{|p{2cm}||X|}
\hline
\textbf{Information}&\textbf{Representation}\\
\hline
\RaggedRight
Detected car count  &Number of detected vehicles in each approach (normalized by maximum capacity of the lane)\\
\hline
Distance & Distance to nearest detected vehicle on each approach\\
\hline
Phase time  & How much time elapsed in current phase (in seconds)\\
\hline
Amber phase  & Indicator of amber phase; 1 if currently in amber phase, otherwise 0 \\
\hline
Current phase & An integer to represent current traffic signal phase\\
\hline
\RaggedRight
Current time (optional) & Current time of the day\\
\hline
\end{tabularx}
    
\end{table}

\subsection{Implementation Strategy}
\label{ss:implement}
There overall implementation strategy of the RL agent includes two phases:
\begin{enumerate}
    \item  \emph{Training Phase:} Train the traffic control agent in a simulator.
    \item \emph{Deployment Phase:} After training, deploy the agent to the real world.
\end{enumerate}
The agent is trained at first on a simulated environment. Thus, the foremost challenge is for the agent to update its weights after deployment to adapt to the changing environment in a way that avoids instability issues that would be problematic in a live environment.

\section{Simulation and Performance}

\subsection{Simulation Settings}
\label{ss:simulationSetting}


We implemented an OpenAI Gym compatible traffic environment \cite{brockman2016openai} -- Gym TrafficLight \cite{gym_trafficlight}. Our simulation is based on 'simple' environments where each approach to the intersection has one lane and the traffic light only has two traffic phases (excluding transition phases). We implemented three different realistic car flow scenarios, denoted as 'simple-sparse', 'simple-medium' and 'simple-dense':
\begin{enumerate}
    \item \textit{Sparse}: Only very few vehicles come to the intersection, which corresponds to a midnight situation. In LuST environment, we use the car flow of 2 AM. 
    \item \textit{Dense}: Many vehicles come to the intersection, which corresponds to the rush hour situation. In LuST environment, we use the car flow of 8 AM.
    \item \textit{Medium}: Intermediate car flow, which corresponds to regular hour traffic. In LuST environment, we use the car flow of 2 PM.
\end{enumerate}

\subsection{Performance in Different Environments}
\label{ss:envComp}
\begin{figure}[ht]
     \centering
     \begin{subfigure}[b]{0.8\linewidth}
         \centering
         \includegraphics[width=\linewidth]{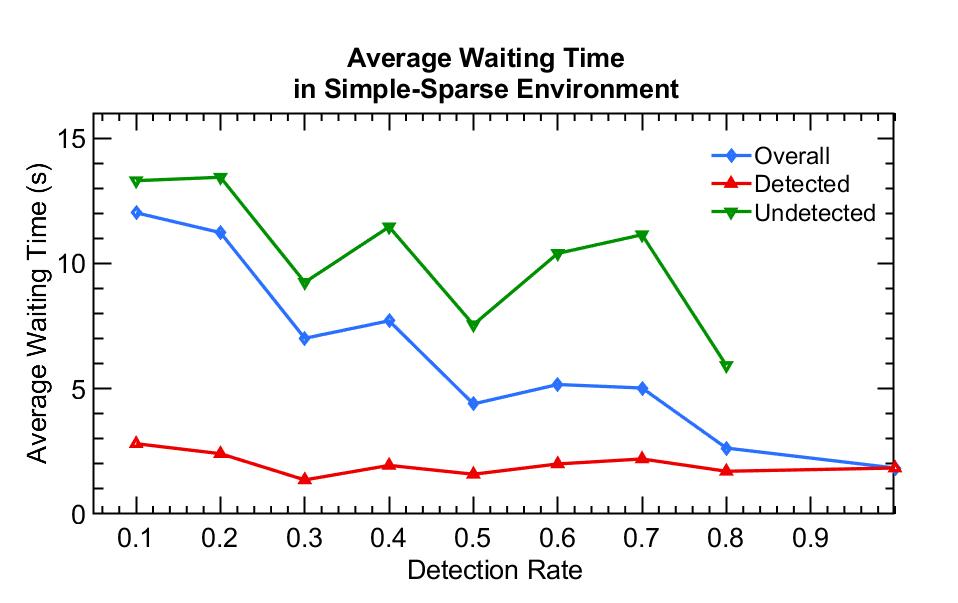}
         \caption{Performance in Simple-sparse environment}
         \label{fig:env_sim_sparse}
     \end{subfigure}
     \hfill
     \begin{subfigure}[b]{0.8\linewidth}
         \centering
         \includegraphics[width=\linewidth]{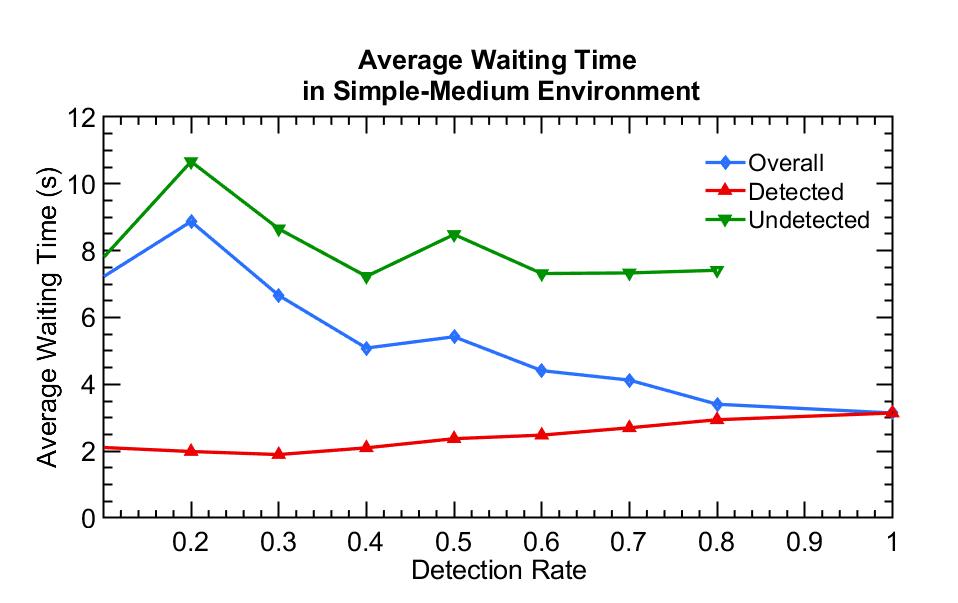}
         \caption{Performance in Simple-medium environment}
         \label{fig:env_sim_med}
     \end{subfigure}
     \begin{subfigure}[b]{0.8\linewidth}
         \centering
         \includegraphics[width=\linewidth]{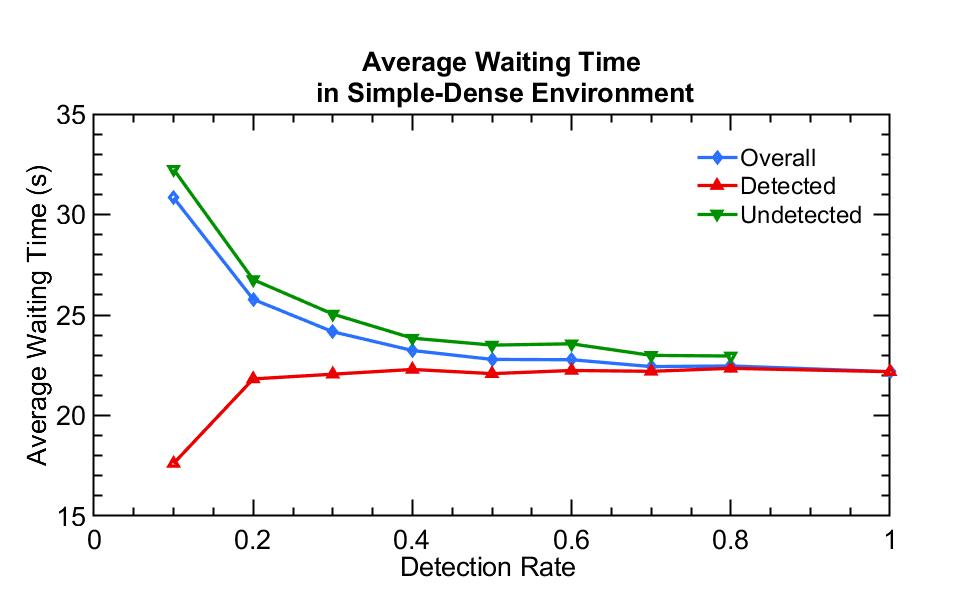}
         \caption{Performance in Simple-dense environment}
         \label{fig:env_sim_dense}
     \end{subfigure}
        \caption{Performance of algorithm for Simple environment under different car flows}
        \label{fig:envComp}
\end{figure}

Figure \ref{fig:envComp} shows the performance found by PPO in different environments, with partial-reward setting.  In all situations, the average waiting time of all vehicles asymptotically decreases as the detection rate increases. This is an ideal property as we want the traffic system to improve its performance when the detection rate gradually increases, as theory in Figure \ref{fig_ill} suggests.

We can also see from the figures that in every case detected vehicles have lower waiting times. More specifically, in the dense traffic flow scenario, the waiting time of detected vehicles rapidly approaches the one of undetected vehicles. This shows that optimizing only the waiting time for detected vehicles, for a dense traffic flow, is equivalent to optimizing the waiting time for all vehicles. This finding is both useful and interesting, as it provides evidence that ITS does not require detecting all vehicles, but only a subset of vehicles, under dense car flow scenarios.

The results agree with the previous finding with Q-learning, published in \cite{zhang2018partially}, suggesting that PPO (and other policy-gradient based algorithms) are able to converge to the same near-optimal solution as Q-learning does for optimizing PD-ITSC systems.

\subsection{Adaptation to Environment Change}
\label{ss:adaptation}
\label{ss:sim_dyn_penn}

In this simulation scenario, we deployed the agents in an environment where the detection rate linearly increased from 0.1 to 1 in 3 years, a more abrupt evolution than we had estimated. We choose this fast evolving environment to reveal an upper bound on the adaptive capacity of each algorithm. 

\begin{figure}[ht]
     \centering
     \begin{subfigure}[b]{0.8\linewidth}
         \centering
         \includegraphics[width=\linewidth]{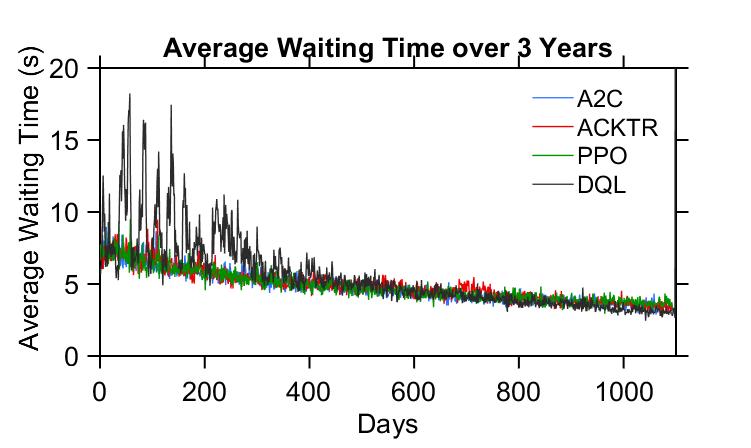}
         \label{fig:dyn_penn_ori}
     \end{subfigure}
     \hfill
        \caption{The average waiting time of all vehicles over 3 years, with detection rate increase linearly from 0.1 to 1}
        \label{fig:dyn_penn_sim}
\end{figure}

Figure \ref{fig:dyn_penn_sim} shows the performance of different agents over 3 years. We see that, in general, the average waiting time of vehicles decreases over these years. However, in the DQL case, the agent sometimes executes a catastrophic update, especially in the early stages of the deployment when detection rate is low. This is due to the high degree of randomness of the environment, and of detected vehicle patterns. This is obviously not a desirable feature, as it can cause major traffic congestion in a real world environment. On the other hand, the three policy-based algorithms show much better stability. Figure \ref{fig:dyn_penn_sim} shows the performance of the three policy-based algorithms. Observe that while the three algorithms indeed perform very similarly in the gradually changing environment,  PPO and A2C outperform ACKTR slightly as they have less deviation.

The results confirm our intuition, as A2C, ACKTR and PPO algorithms are based on policy gradient method, which, at each update, try to improve their policy based on their approximation of policy gradient, and hence the performance after each iteration gets better; however, Q-learning method is value-based, and rapid Q-value updates in a noisy environment sometimes lead to a 'bad update'.

\section{Discussion}
Preliminary results reported in this paper have shown that in the deployment phase, Q-learning does not safely adapt itself to a dynamic detection rate in a given environment. However, all three policy-gradient algorithms, A2C, ACKTR, and PPO, are practical solutions in satisfying the objectives considered in this paper. Further research is needed to evaluate other important aspects, such as the impact of rewards and the application to more complex environments. We plan to report the results of that investigation in a future paper. 



\section{Conclusion}
In this paper, we investigated several RL algorithms to solve the adaptation problem in partially detected intelligent traffic signal control systems. We proposed to solve this problem by using partial reward and policy-based algorithms such as A2C, ACKTR, and PPO.

Simulation results showed that, 
with the methods proposed, the system preserves all the profitable properties of PD-ITSC systems. 
The detected vehicles always have less waiting time than undetected vehicles, but this gap becomes smaller with increased detection rates and  increased car flow rates.


\bibliographystyle{IEEEtran}
\bibliography{sigproc}  
\end{document}